\documentclass[preprint,showpacs,preprintnumbers,amsmath,amssymb]{revtex4}
\usepackage{graphicx}
\usepackage{dcolumn}
\usepackage{bm}

\newcommand{\be}{\begin{equation}}
\newcommand{\en}{\end{equation}}
\newcommand{\bea}{\begin{eqnarray}}
\newcommand{\ena}{\end{eqnarray}}

\begin{document}


\title{ Tachyon-Chaplygin inflationary universe model  }

\author{Sergio del Campo}
 \email{sdelcamp@ucv.cl}
\affiliation{ Instituto de F\'{\i}sica, Pontificia Universidad
Cat\'{o}lica de Valpara\'{\i}so, Casilla 4059, Valpara\'{\i}so,
Chile.}
\author{Ram\'on Herrera}
\email{ramon.herrera@ucv.cl} \affiliation{ Instituto de
F\'{\i}sica, Pontificia Universidad Cat\'{o}lica de
Valpara\'{\i}so, Casilla 4059, Valpara\'{\i}so, Chile.}

\date{\today}

\begin{abstract}
 Tachyonic inflationary universe model in the context of a  Chaplygin
 gas equation of state is studied. General
 conditions  for this model to be realizable are
 discussed. By using an effective exponential potential we
 describe in great details the characteristic of the inflationary
 universe model. The parameters of the model are restricted by
 using recent astronomical  observations.
\end{abstract}

\maketitle

\section{Introduction}

It is well known that inflation is to date the most compelling
solution to many long-standing problems of the Big Bang model
(horizon, flatness, monopoles, etc.) \cite{guth,infla}. One of the
success of the inflationary universe model is that it provides a
causal interpretation of the origin of the observed anisotropy of
the cosmic microwave background (CMB) radiation, and also the
distribution of large scale structures \cite{astro}.

In concern to higher dimensional theories, implications of
string/M-theory to Friedmann-Robertson-Walker (FRW) cosmological
models have recently attracted  great deal of attention, in
particular, those related to brane-antibrane configurations such
as space-like branes\cite{sen1}. In recent times a great amount of
work has been invested in studying the inflationary model with a
tachyon field. The tachyon field associated with unstable D-branes
might be responsible for cosmological inflation in the early
evolution of the universe, due to tachyon condensation near the
top of the effective scalar potential \cite{sen2}, which could
 also add some new form of cosmological dark matter at late times
\cite{Sami_taq}. In fact, historically, as was empathized by
Gibbons \cite{gibbons}, if the tachyon condensate starts to roll
down the potential with small initial velocity, then a universe
dominated by this new form of matter will smoothly evolve from a
phase of accelerated expansion (inflation) to an era dominated by
a non-relativistic fluid, which could contribute to the dark
matter detected in these days.

On the other hand, the generalized Chaplygin gas    has been
proposed as an  alternative model for  describing the accelerating
of the universe. The generalized Chaplygin gas is described by an
exotic equation of state of the form \cite{Bento}
\begin{equation}
 p_{ch} = - \frac{A}{\rho_{ch}^\beta},\label{1}
\end{equation}
where $\rho_{ch}$ and $p_{ch}$ are the energy density and pressure
of the generalized Chaplygin gas, respectively. $\beta$ is a
constant that lies in  the range $0 <\beta\leq 1$, and $A$ is a
positive constant. The original  Chaplygin gas corresponds to the
case $\beta = 1$ \cite{2}. Inserting this equation of state into
the relativistic energy conservation equation leads to an energy
density given by \cite{Bento}
\begin{equation}
 \rho_{ch}=\left[A+\frac{B}{a^{3(1+\beta)}}\right]^{\frac{1}{1+\beta}},
 \label{2}
\end{equation}
 where $a$ is the
scale factor  and $B$ is a positive integration constant.

The Chaplygin gas emerges as a effective fluid of a generalized
d-brane in a (d+1, 1) space time, where the   action can be
written as a generalized Born-Infeld action \cite{Bento}. These
models have been extensively studied in the literature
\cite{other}. The model parameters were constrained using currents
cosmological observations, such as, CMB \cite{CMB} and  supernova
of type Ia (SNIa) \cite{SIa}.

In the  model of  Chaplygin inspired inflation usually the scalar
field, which drives inflation, is the standard inflaton field,
where the energy density given by Eq.(\ref{2}), can be extrapolate
for  obtaining a successful inflation period with a Chaplygin gas
model\cite{Ic}. Recently, the dynamics of the early universe and
the initial conditions for inflation in a model with radiation and
a Chaplygin gas was studied  in Ref.\cite{Monerat:2007ud}.  As far
as we know, a Chaplygin inspired inflationary model in which a
tachyonic field is considered has not been  studied. The main goal
of the present work is to investigate the possible realization of
a Chaplygin inflationary universe model, where the energy density
is driven by a tachyonic field.  We use  astronomical data for
constraining the parameters appearing in this model.

The outline of the paper is a follows. The next section presents a
short review of the modified Friedmann equation by using a
Chaplygin gas, and  we present the tachyon-Chaplygin inflationary
model. Section \ref{sectpert} deals with the calculations of
cosmological perturbations in general term.  In Section
\ref{exemple} we use an exponential potential for obtaining
explicit expression for the model. Finally, Sect.\ref{conclu}
summarizes our findings.

\section{The modified Friedmann equation and the Tachyon-Chaplygin Inflationary phase. }

We start by writing down  the modified Friedmann equation, by
using the FRW metric. In particular, we assume that the
gravitational dynamics  give rise to a modified Friedmann
equation of form
\begin{equation}
H^2=\kappa\left[A+\rho_\phi^{(1+\beta)}\right]^{\frac{1}{1+\beta}},
\label{HC}
\end{equation}
where $\kappa=8\pi G/3=8\pi/3m_p^2$ (here $m_p$ represents the
Planck mass), $\rho_\phi=\frac{V(\phi)}{\sqrt{1-\dot{\phi}^2}}$,
  and $V(\phi)=V$ is
the scalar tachyonic   potential.  The modification is realized
from an extrapolation of Eq.(\ref{2}), where  the density matter
$\rho_m\sim a^{-3}$ in introduced in such  a way that we may write
\begin{equation}
 \rho_{ch}=\left[A+\rho_m^{(1+\beta)}\right]^{\frac{1}{1+\beta}},
 \label{extr}
\end{equation}
and  thus, we identifying $\rho_m$ with the contributions of the
tachyon field for give Eq.(\ref{HC}). The generalized Chaplygin
gas model may be viewed as a modification of gravity, as described
in Ref.\cite{Ber}, and for chaotic inflation, in Ref.\cite{Ic}.
Different modifications of gravity have been proposed in the last
few years, and there has been a lot of interest in the
construction of early universe scenarios in higher-dimensional
models motivated by string/M-theory \cite{Ran}. It is  well-known
that these modifications can lead to important changes in the
early universe. In the following  we will take $\beta=1$ for
simplicity, which means the usual Chaplygin gas.

From Eq.(\ref{HC}), the dynamics of the cosmological model in the
tachyon-Chaplygin inflationary scenario is described by the
equations
\begin{equation}
H^2=\kappa\,\sqrt{A+\rho_\phi^2}=
\kappa\,\sqrt{A+\frac{V^2}{1-\dot{\phi}^2}}\label{mm3},
\end{equation}
and

 \be \frac{\ddot{\phi}}{1-\dot{\phi}^2}+\,3H \;
\dot{\phi}+\frac{V'}{V}=0, \label{key_01}
 \en
where dots mean derivatives with respect to the cosmological time
and
 $V'=\partial V(\phi)/\partial\phi$. For convenience we will use
 units in which $c=\hbar=1$.

During the inflationary epoch the energy density associated to the
 tachyon field is of the order of the potential, i.e.
$\rho_\phi\sim V$. Assuming the set of slow-roll conditions, i.e.
$\dot{\phi}^2 \ll 1$ and $\ddot{\phi}\ll 3H\dot{\phi}$
\cite{gibbons,Fairbairn:2002yp}, the Friedmann equation
(\ref{mm3})  reduces  to
\begin{eqnarray}
H^2=\kappa\,\sqrt{A+\rho_\phi^2}\approx\kappa\,\sqrt{A+V^2},\label{inf2}
\end{eqnarray}
and  Eq. (\ref{key_01}) becomes
\begin{equation}
3H \dot{\phi}\approx-\frac{V'}{V}. \label{inf3}
\end{equation}

Introducing the dimensionless slow-roll parameters
\cite{Hwang:2002fp}, we write
\begin{equation}
\varepsilon=-\frac{\dot{H}}{H^2}\simeq\frac{1}{6\kappa}\,\frac{V'^2}{(A+V^2)^{3/2}},\label{ep}
\end{equation}

\begin{equation}
\eta=-\frac{\ddot{\phi}}{H\,
\dot{\phi}}\simeq\,\frac{1}{3\kappa\sqrt{A+V^2}}\left[\frac{V''}{V}-\frac{V'^2}{V^2}
-\frac{1}{2}\frac{V'^2}{(A+V^2)}\right]\label{eta},
\end{equation}
and
\begin{equation}
\gamma=-\frac{V'\,\dot{\phi}}{2\,H\,V}\simeq\frac{1}{6\kappa}\,\frac{V'^2}{V^2\,(A+V^2)^{1/2}}.\label{ep}
\end{equation}

The condition under  which inflation  takes place can be
summarized with the parameter $\varepsilon$ satisfying  the
inequality $\varepsilon<1.$, which  is analogue to the requirement
that  $\ddot{a}> 0$. This condition could be written in terms of
the  tachyon potential and its  derivative $V'$, which becomes
\begin{equation}
(A+V^2)^{3/2}> \frac{V'^2}{6\kappa}.\label{cond}
\end{equation}

Inflation ends when the universe heats up at a time when
$\varepsilon\simeq 1$, which implies
\begin{equation}
(A+V_f^2)^{3/2}\simeq \frac{V_f'\,^2}{6\kappa}.
\end{equation}
The number of e-folds at the end of inflation is given by
\begin{equation}
N=-3\;\kappa\,\int_{\phi_{*}}^{\phi_f}\frac{\sqrt{A+V^2}}{V'}
\,V\,d\phi',\label{N}
\end{equation}
or equivalently
\begin{equation}
N=-3\;\kappa\,\int_{V_{*}}^{V_f}\frac{\sqrt{A+V^2}}{V'^2}
\,V\,d\,V.\label{NV}
\end{equation}

  In the following, the subscripts  $*$ and $f$ are
used to denote  the epoch when the cosmological scales exit the
horizon and the end of  inflation, respectively.

\section{Perturbations\label{sectpert}}

In this section we will study the scalar and tensor perturbations
for our model. The general perturbed metric about the flat FRW
background \cite{Bar} is :
\begin{equation}
ds^2=-(1+2A)dt^2+2a(t)B_{,\,i}dx^{i}dt+a(t)^2[(1-2\psi)\delta_{ij}+2E_{,\,i,\,j}+2h_{ij}]dx^idx^j\,,
\end{equation}
where $A$, $B$, $\psi$ and $E$ correspond to the scalar-type
metric perturbations, and $h_{ij}$ characterizes the
transverse-traceless tensor-type perturbation. We introduce
comoving curvature perturbations,
$\cal{R}=\psi+H\delta\phi/\dot{\phi}$, where $\delta\phi$ is the
perturbation of the scalar field $\phi$.  For a  tachyon field the
power spectrum of the curvature perturbations  is given  in the
slow-roll approximation by following expression
\cite{Hwang:2002fp}

\begin{equation}
{\cal{P}_R}\simeq\left(\frac{H^2}{2\pi\dot{\phi}}\right)^2\,\frac{1}{V}\simeq
\frac{9\kappa^3}{4\pi^2}\,\left[\frac{V\,(A+V^2)^{3/2}}{V'^2}\right].\label{dp}
\end{equation}

The scalar spectral index $n_s$ is given by $ n_s -1 =\frac{d
\ln\,{\cal{P}_R}}{d \ln k}$,  where the interval in wave number is
related to the number of e-folds by the relation $d \ln k(\phi)=-d
N(\phi)$. From Eq.(\ref{dp}), we get, by using the slow-roll
parameters,
\begin{equation}
n_s  \approx\, 1\,-2(2\varepsilon-\eta-\gamma),\label{ns1}
\end{equation}
 or equivalently
\begin{equation}
n_s  \approx\, 1\,-\frac{1}{\kappa\,(A+V^2)^{1/2}}\,\left[
\frac{V'^2}{(A+V^2)}+\frac{V'^2}{3\,V^2}-\frac{2}{3}\frac{V''}{V}\right].\label{nsa}
\end{equation}

Note that in the limit $A\rightarrow 0$, the scalar spectral index
$n_s$  coincides with that corresponding to a single tachyon field
\cite{Sami_taq}.

One of the interesting features of the three-year data set from
Wilkinson Microwave Anisotropy Probe (WMAP) is that it hints at a
significant running in the scalar spectral index $dn_s/d\ln
k=\alpha_s$ \cite{astro}. From Eq.(\ref{ns1}) we get  that the
running of the scalar spectral index becomes

\begin{equation}
\alpha_s=\left(\frac{4\,(A+V^2)}{V\;V'}\right)\,[2\varepsilon_{,\,\phi}-\eta_{,\,\phi}-\gamma_{,\,\phi}]
\;\varepsilon.\label{dnsdk}
\end{equation}
In models with only scalar fluctuations the marginalized value for
the derivative of the spectral index is approximately $-0.05$ from
WMAP-three year data only \cite{astro}.

On the other hand, the generation of tensor perturbations during
inflation would produce  gravitational waves and its amplitudes
are given by \cite{Bar}
\begin{equation}
{\cal{P}}_g=24\kappa\,\left(\frac{H}{2\pi}\right)^2
\simeq\frac{6}{\pi^2}\,\kappa^2\,(A+V^2)^{1/2},\label{ag}
\end{equation}
where the spectral index $n_g$ is given by $
n_g=\frac{d{\cal{P}}_g}{d\,\ln k}=-2\,\varepsilon$.

From expressions (\ref{dp}) and (\ref{ag}) we write  the
tensor-scalar ratio as
\begin{equation}
R(k)=\left.\left(\frac{{\cal{P}}_g}{P_{\cal
R}}\right)\right|_{k=k_*}
\simeq\left.\left(\frac{24\kappa\,V\,\dot{\phi}^2}{H^2}\right)\right|_{\,k=k_*}=
\left.\left(\frac{8}{3\,\kappa}\,\frac{V'^2}{V\;(A+V^2)}\right)\right|_{\,k=k_*}.
\label{Rk}\end{equation} Here, $k_*$  is referred to $k=Ha$, the
value when the universe scale  crosses the Hubble horizon  during
inflation. Note that the  consistency relation for the
tensor-scalar ratio, $R=-8\,n_g$, becomes similar to that
corresponding to the standard scalar field \cite{Steer:2003yu}.

Combining  WMAP three-year data\cite{astro} with the Sloan Digital
Sky Survey  (SDSS) large scale structure surveys \cite{Teg}, it is
found an upper bound for $R$ given by $R(k_*\simeq$ 0.002
Mpc$^{-1}$)$ <0.28\, (95\% CL)$, where $k_*\simeq$0.002 Mpc$^{-1}$
corresponds to $l=\tau_0 k\simeq 30$,  with the distance to the
decoupling surface $\tau_0$= 14,400 Mpc. The SDSS  measures galaxy
distributions at red-shifts $a\sim 0.1$ and probes $k$ in the
range 0.016 $h$ Mpc$^{-1}$$<k<$0.011 $h$ Mpc$^{-1}$. The recent
WMAP three-year results give the values for the scalar curvature
spectrum $P_{\cal R}(k_*)\simeq 2.3\times\,10^{-9}$ and the
scalar-tensor ratio $R(k_*)=0.095$. We will make use of these
values  to set constrains on the parameters appearing in  our
model.

\section{Exponential potential in a Tachyon-Chaplygin gas. \label{exemple}}
Let us consider a  tachyonic effective potential $V(\phi)$,  with
the properties satisfying   $V(\phi)\longrightarrow$ 0 as
$\phi\longrightarrow \infty$. The exact form of the potential is
$V(\phi)=\left ( 1+\alpha \phi \right )\exp(-\alpha \phi)$, which
in the case when $\alpha \rightarrow 0$, we may use the asymptotic
exponential expression. This form for the potential is derived
from string theory calculations\cite{ku,sen2}. Therefore, we
simple use
\begin{equation}
V(\phi)=V_0 e^{-\alpha\phi},\label{pot}
\end{equation}
where $\alpha$ and $V_0$ are free parameters. In the following we
will restrict ourselves to the case in which  $\alpha> 0$. Note
that $\alpha$ represents the tachyon mass
\cite{Fairbairn:2002yp,delaMacorra:2006tm}. In Ref.\cite{Sami_taq}
is given an estimation of these parameters  in the limit
$A\rightarrow 0$. Here, it was found  $V_0\sim 10^{-10}m_p^4$ and
$\alpha\sim 10^{-6} m_p$. We should mention here that the caustic
problem  with multi-valued regions for scalar Born-Infeld theories
with an exponential potential results in high order spatial
derivatives of the tachyon field, $\phi$, become
divergent\cite{ko}.

From Eq.(\ref{NV}) the number of e-folds results in
\begin{equation}
N=\frac{3\kappa}{\alpha^2}\;[h(V_f)-h(V_*)],
\end{equation}
where
\begin{equation}
h(V)=\left(\sqrt{A}\;\ln\left[\frac{2(\sqrt{A}+W)}{A\,V}\right]-W\right),
\end{equation}
and $W=W(V)=(A+V^2)^{1/2}$.

On the other hand, we may establish that the end of  inflation is
governed by the condition  $\varepsilon=1$, from which we get that
the square of the scalar tachyon  potential becomes
\begin{equation}
V^2(\phi=\phi_f)=V_f^2=\frac{1}{108\kappa^2}\,\left[\alpha^4-108\,A\,\kappa^2+
\frac{\alpha^8-216\,A\,\kappa^2\,\alpha^4}{\Im}+\Im\right].
\end{equation}
Here, $\Im$ is given by
$$
\Im=\alpha^{4/3}\;[\alpha^{8}-324\,A\kappa^2\alpha^4+17496\,A^2\,\kappa^4\,+648\sqrt{3}\,A^{3/2}\kappa^3\,
\sqrt{243A\kappa^2-\alpha^4}\,]^{1/3}.
$$
Note that in the limit $A\rightarrow 0$ we obtain
$V_f=\frac{\alpha^2}{6\kappa}$, which  coincides with that
reported in Ref.\cite{Sami_taq}.

From Eq.(\ref{dp}) we obtain that the scalar power spectrum is
given by
\begin{equation}
P_{\cal R}(k)\approx\left.
\;\frac{9\kappa^3}{4\pi^2\,\alpha^2}\left[\frac{(A+V^2)^{3/2}}{V}\right]\right|_{\,k=k_*},\label{ppp}
\end{equation}
and from Eq.(\ref{Rk}) the tensor-scalar ratio becomes
\begin{equation}
R(k)\approx\;\left.\frac{8}{3\kappa}\left[\frac{\alpha^2\;V}{(A+V^2)}\right]\right|_{\,k=k_*}.\label{rrrr}
\end{equation}

By using the WMAP three year data where $P_{\cal R}(k_*)\simeq
2.3\times 10^{-9}$ and $R(k_*)=0.095$,  we obtained from
Eqs.(\ref{ppp}) and (\ref{rrrr}) that
\begin{equation}
A\simeq\,\frac{10^{-19}}{\kappa^4}\left[1-\frac{2\times10^{-22}}{\kappa^2\,\alpha^4}\right]\;,
\label{A}
\end{equation}
and
\begin{equation}
V_*\simeq\,\frac{5\times10^{-21}}{\alpha^2\,\kappa^3}.\label{constrain}
\end{equation}
From Eq.(\ref{A}) and  since $A>0$,  $\alpha$ satisfies  the
inequality $\alpha>10^{-6}m_p$. This inequality allows us to
obtain an upper limit for the tachyon potential $V(\phi)$ evaluate
when the cosmological scales exit the horizon, i.e.
$\kappa^2\,V_*<5\times10^{-10}$. Note that in the limit
$A\rightarrow 0$,  the constrains $\alpha\sim 10^{-6} m_p$ and
$V_*\sim10^{-11}m_p^4$ are recovered \cite{Sami_taq}.

By using an exponential potential we obtain from Eq.(\ref{nsa})
\begin{equation}
\frac{(2V^2-A)^2}{(V^2+A)^3}=\frac{9\kappa^2}{\alpha^4}\,(n_s-1)^2.
\end{equation}
This expression  has roots that can be solved analytically for the
tachyonic potential  $V$, as a function of $n_s$, $A$ and
$\alpha$.  For a real root solution for $V$,  and from
Eq.(\ref{A}) and (\ref{dnsdk})   we  obtain a relation of the form
$\alpha_s=f(n_s)$ for a fixed value  of $\alpha$. In Fig. 1 we
have plotted  the running spectral index $\alpha_s$ versus the
scalar spectrum index $n_s$. In doing this, we have taken two
different values for the parameter $\alpha$. Note that for
$\alpha>10^{-5}m_p$  and for a given $n_s$ the values of
$\alpha_s$ becomes far from that registered  by  the WMAP
three-year data. For example, for $\alpha=10^{-4}m_p$ and
$n_s=0.97$ we obtained that $\alpha_s\simeq-7\times10^{4}$. Note
also that from Fig. 1  the WMAP-three data favors the parameter
$\alpha$ lies in the range $10^{-6}<\alpha/m_p\lesssim10^{-5}$.
The lower limit for $\alpha$ results by considering   $A>0$ and
its upper limit from the relation  $\alpha_s=f(n_s)$ (see Fig. 1).
In example, for $\alpha=4\times 10^{-6} m_p$ and $n_s=0.97$ we get
 the values $A\simeq 2\times10^{-23}m_p^8$, $V_*\simeq
5\times 10^{-13}m_p^4$ and $\alpha_s\simeq -0.02$. Also,  the
number of e-folds, $N$,  becomes of the order of $N\sim 41$. This
lower value  is not a problem since, in the context of the
tachyonic curvaton reheating, the e-folding could be of the order
of  $40$ or $50$, due to the inflationary scale is lower
\cite{Campuzano:2005qw}. We  should note also that the $A$
parameter becomes smaller  by  fourth order of magnitude when it
is compared with the case of Chaplygin inflation with a  standard
scalar field  \cite{Ic}.

Of particular interest is the quantity known as the reheating
temperature. The reheating temperature is associated to the
temperature of the universe when the Big Bang scenario begins (
the radiation epoch). In general, this epoch is generated by the
decay of the inflaton field which leads to  creation of particles
of different kinds\cite{Abb}. The stage of oscillations of the
scalar field is a essential part for the standard mechanism of
reheating \cite{BB}. However, this mechanic does not work when the
inflaton potential does not have a minimum \cite{K}. These models
are known in the literature like non oscillating models, or simply
NO models \cite{7,8}. An alternative mechanism of reheating in NO
models is the introduction of the curvaton field \cite{14}. In the
following, let us brief comment on this and  we  give an
estimation of the reheating temperature for our model. We follow a
similar procedure described in Refs.\cite{Campuzano:2005qw} and
\cite{RH}.

In the context of the curvaton scenario, reheating does occur at
the time when the curvaton decays, but only in the period when the
curvaton dominates. In contrast, if the curvaton decays before its
density dominates the universe, reheating occurs when the
radiation due to the curvaton decay manages to dominate the
universe. During the epoch in which the curvaton decays after it
dominates it is found  that the reheating temperature, $T_{rh}$,
is of the order of $T_{rh}\sim 10^{-10} m_p$. Here, we have used
that the curvaton field $\sigma_*$ becomes the order of
$\sigma_*\sim m_p$, $A=2\times10^{-23}m_p^8$, $V_*\simeq 5\times
10^{-13}m_p^4$ and from Eq.(\ref{inf2}),  $H_*\sim 10^{-14} GeV$.
We should note that this value for $T_{rh}$ could be modified, if
the decay of the curvaton field happens after domination (see
Ref.\cite{RH}).

\begin{figure}[th]
\includegraphics[width=4.0in,angle=0,clip=true]{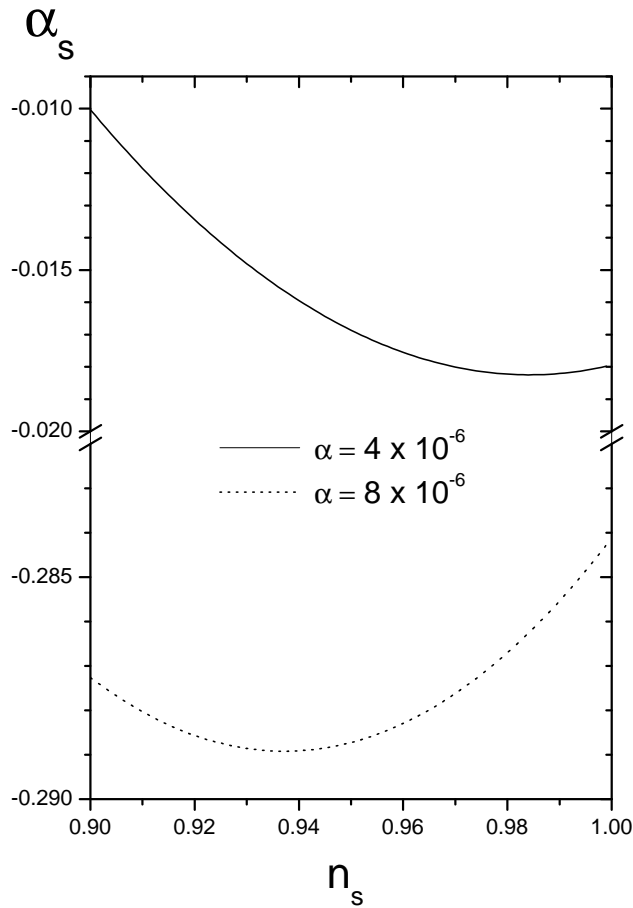}
\caption{Evolution of the  running  scalar spectral index
$\alpha_s$ versus the scalar spectrum index $n_s$, for two
different values of the parameter $\alpha$.
 \label{rons}}
\end{figure}


\section{Conclusions \label{conclu}}

In this paper we have studied the tachyon-Chaplygin inflationary
model. In the slow-roll approximation we have found a general
relation between the tachyonic potential and its derivative. This
has led us to a general criterium for inflation to occur (see
Eq.(\ref{cond})). We  have also obtained   explicit expressions
for the corresponding scalar spectrum index $n_s$ and its running
$\alpha_s$.

By using  an exponential potential with $\alpha$ fixed (see
Eq.(\ref{pot})) and from the WMAP three year data,  we  found the
values of the parameter $A$ and an upper limit for the tachyon
potential $V_*$. In order to bring  some explicit results we have
taken  $\alpha=4\times 10^{-6} m_p$ and $n_s=0.97$, from which we
get the values $A\simeq 2\times10^{-23}m_p^8$, $V_*\simeq 5\times
10^{-13}m_p^4$ and $\alpha_s\simeq -0.02$. The restrictions
imposed  by currents  observational data allowed us to establish a
small range for the parameter $\alpha$, which become
$10^{-6}<\alpha/m_p\lesssim 10^{-5}$. From this range,  and from
Eqs.(\ref{A}) and (\ref{constrain}), we obtained the ranges
$0<A/m_p^8\lesssim 10^{-23}$ and $8.5\times10^{-14}\lesssim
V_*/m_p^4<8.5\times10^{-12}$.

In the context of the curvaton scenario, we gave an estimation of
the reheating temperature, when the curvaton decay occurs  after
it dominates. However, a more accurate calculation for the
reheating temperature $T_{rh}$ in the curvaton scenario, would  be
necessary for establishing  some constrains on other parameters
appearing  in our model. We hope to return to this point in the
near future.

\begin{acknowledgments}
 S. del C. was supported from COMISION NACIONAL DE CIENCIAS Y
TECNOLOGIA through FONDECYT Grant N$^{0s}$.  1040624, 1051086 and
1070306 and also it was partially supported by PUCV Grant N$^0$.
123.787/2007. R.H. was supported by the ``Programa Bicentenario de
Ciencia y Tecnolog\'{\i}a" through the Grant ``Inserci\'on de
Investigadores Postdoctorales en la Academia" \mbox {N$^0$
PSD/06}.
\end{acknowledgments}


\end{document}